\documentclass[twocolumn,showpacs,preprintnumbers,superscriptaddress]{revtex4}
\usepackage{}

\usepackage{amsmath,amssymb,graphicx,bm}

\begin{document}

\title{Test of conformal gravity with astrophysical observations}

\author{Rongjia Yang}
\email{yangrongjia@tsinghua.org.cn}
 \affiliation{College of Physical Science and Technology, Hebei University, Baoding 071002, China}
 \affiliation{Department of Physics, Tsinghua University, Beijing 100084, China}
\author{Bohai Chen}
 \affiliation{College of Physical Science and Technology, Hebei University, Baoding 071002, China}

\author{Haijun Zhao}
 \affiliation{School of Physics and Information Science, Shanxi Normal University, Linfen 041004, China}

\author{Jun Li}
 \affiliation{College of Physical Science and Technology, Hebei University, Baoding 071002, China}

\author{Yuan Liu}
 \affiliation{Key Laboratory of Particle Astrophysics, Institute of High Energy Physics, Chinese Academy of Sciences, P.O.Box 918-3, Beijing 100049, China}


\begin{abstract}
Since it can describe the rotation curves of galaxies without dark matter and can give rise to accelerated expansion, conformal gravity attracts much attention recently. As a theory of modified gravity, it is important to test conformal gravity with astrophysical observations. Here we constrain conformal gravity with SNIa and Hubble parameter data and investigate whether it suffers from an age problem with the age of APM~08279+5255. We find conformal gravity can accommodate the age of APM~08279+5255 at 3 $\sigma$ deviation, unlike most of dark energy models which suffer from an age problem.

\end{abstract}

\maketitle

\section{Introduction}
Many astronomical observations indicate that the Universe is undergoing late-time acceleration. An unknown energy component, dubbed as dark energy, is usually proposed to explain the accelerated expansion.
The simplest and most attractive candidate is the
cosmological constant model ($\Lambda$CDM). This model is consistent
with most of current astronomical observations, but suffers from the cosmological constant problem \cite{weinberg}, as well as age problem \cite{Yang2010}.
It is thus natural to pursue alternative possibilities to explain
the mystery of the accelerated expansion. Over the past numerous dark energy models have been proposed, such as
quintessence, phantom, k-essence, quintom, tachyon, etc. Rather than by introducing a dark energy, modified gravity, such as $f(R)$ theories (for reviews, see e. g. \cite{Felice, Sotiriou, Capozziello,Clifton2012}), $f(T)$ theories (see e. g. \cite{Bengochea,Yang2011}), and $f(R, R_{\mu\nu}R^{\mu\nu})$ \cite{Carroll,Clifton,Navarro} theories,
are proposed as ways to obtain a late accelerated expansion by modifying the Lagrangian of general relativity. Conformal gravity (CG, following the original work by Weyl \cite{Weyl1918}, for reviews, see \cite{Nesbet,Scholz,Mannheim2011}), as a special $f(R, R_{\mu\nu}R^{\mu\nu})$ theory, can give rise to accelerated expansion \cite{Mannheim2001}. It was also claimed that CG can describe the rotation curves of galaxies without dark matter \cite{Mannheim1997}. A static solution for a charged black hole in CG was presented in \cite{Said}. Perturbations in the cosmology associated with the conformal gravity theory were investigated in \cite{Mannheim2012}. It had been shown that currently available SNIa and GRB samples were accommodated well by CG \cite{Diaferio2011}. Although CG attracts much attention, it is also confronted with some challenges. It was argued in \cite{Flanagan} that in the limit of weak fields and slow motions CG does not agree with the predictions of general relativity, and it is therefore ruled out by Solar System observations (Recently, however, it was indicated in \cite{Cattani2013} that conformal gravity can potentially test well against all astrophysical observations to date). CG can not describe the phenomenology of gravitational lensing \cite{Pireaux} and of clusters of galaxies \cite{Horne}. CG is not able to explain the properties of X-ray galaxy clusters without resorting to dark matter \cite{Diaferio}. CG can not pass the primordial nucleosynthesis test \cite{Elizondo}, however, this is an open problem, because all the possible mechanisms producing deuterium are still incomplete.

Besides the dark energy problem, the age problem is another important test for cosmological models. A spatially flat
Friedmann-Robertson-Walker (FRW) universe dominated by matter (with age $T=2/3H_0$), for example, is ruled out unless $h<0.48$ ($h=H/100$ km s$^{-1}$Mpc$^{-1}$) \cite{Pont}, compared with the 14 Gyr age of the Universe inferred from old globular clusters.
Introducing dark energy can not only explain the accelerated expansion, but also reconcile the age problem. However, the discovery of an old quasar APM~08279+5255 at $z=3.91$ which was initially estimated to be around 2-3 Gyr \cite{Komossa2003} and re-evaluated to be 2.1 Gyr \cite{Friaca2005} has once again led to an age problem for cosmological models, such as $\Lambda$CDM \cite{Friaca2005, Alcaniz2003}, the creation of cold dark matter models \cite{Chen2012},
$\Lambda(t)$ model \cite{Cunha2004},
the new agegraphic dark energy \cite{Li2011}, parametrized variable Dark Energy
Models \cite{Barboza2008, Dantas2007}, the $f(R)=\sqrt{R^2-R^2_0}$ model \cite{Movahed2007}, quintessence \cite{Capozziello2007, Jesus2008}, holographic dark energy model \cite{Granda2009,Wei2007}, braneworld modes \cite{Movahed2007a, Pires2006, Movahed2008, Alam2006}, and other models \cite{Sethi2005, Abreu2009, Santos2008}. Most of these researches imposed a priori on the Hubble constant $H_0$, or on the matter
density parameter $\Omega_{\rm m0}$, or on other parameters. To a certain extent, the age problem, is dependent on the values of $H_0$ or $\Omega_{\rm m0}$ one takes. In \cite{Yang2010}, the age problem in $\Lambda$CDM had been investigated in a consistent way, not any special values of $\Omega_{\rm
m0}$ or $H_0$ has been taken with prejudice.

As a theory of modified gravity, it is important to test conformal gravity with astrophysical observations. Here we aim to test CG with observational data including the age of APM~08279+5255. Following \cite{Yang2010}, we obtain directly observational constraints on parameters from SNIa and $H(z)$ data in the framework of the CG, then investigate whether it suffers from an age problem in the parameter space allowed by these observations.

The structure of this paper is as follows. In Section II, we review the cosmology in CG.
In Section III, we consider constraints on the parameters of the cosmology in CG from
SNIa and $H(z)$ data, and use the best-fit values to discuss the ``age
problem''. Conclusions and discussions are given in Section IV.

\section{Cosmology in Conformal gravity}
The action of CG with matter is given by
\begin{eqnarray}
\label{action}
\mathcal{I}=-\alpha_{\rm g} \int C_{\mu\nu\kappa\lambda} C^{\mu\nu\kappa\lambda}\sqrt{-g} {\rm d}^4x+\mathcal{I_{\rm m}},
\end{eqnarray}
where $\alpha_{\rm g}$ is a dimensionless coupling constant and $C_{\mu\nu\kappa\lambda}$ the Weyl tensor. This action is invariant under local conformal transformations: $g_{\mu\nu}\rightarrow e^{2\alpha(x)}g_{\mu\nu}$. This symmetry forbids the presence of any $\Lambda \sqrt{-g}{\rm d}^4x$ term in the action, so CG does not suffer from the cosmological constant problem. Secondly, since $\alpha_{\rm g}$ is dimensionless, unlike general relativity CG is renormalizable \cite{Mannheim2011}. Thirdly, though the equations of motion are fourth-order, CG is a ghost-free theory \cite{Bender2008}. However, CG is also confronted with some challenges as discussed in the previous section. The matter action can be of the form \cite{Elizondo,Mannheim1992}
\begin{eqnarray}
\label{actionm}&&
\mathcal{I_{\rm m}}=-\int \sqrt{-g} {\rm d}^4x \nonumber\\
&&\times \left[\frac{1}{2}S^{;\mu}S_{;\mu}+\lambda S^4-\frac{1}{12}S^2R+i\overline{\psi}\gamma^\mu D_\mu \psi-\zeta S\overline{\psi}\psi\right],
\end{eqnarray}
where scalar field $S(x)$ is introduced to spontaneously break the conformal symmetry and renders the particles massive, $\psi$ is a fermion field representing all matter field, $D_\mu=\partial_\mu + \Gamma_\mu$ is the covariant derivative with $\Gamma_\mu$ the fermion spin connection, $\lambda$ and $\zeta$ are dimensionless coupling constants, $\gamma^\mu$ are the general relativistic Dirac matrices. $\lambda S^4$ represents the negative minimum of the Ginzburg-Landau potential \cite{Mannheim2001} with $\lambda<0$. For action (\ref{action}), variation with respect to the metric generates the field equations
\begin{eqnarray}
\label{field}
4\alpha_{\rm g}W_{\mu\nu}=T_{\mu\nu},
\end{eqnarray}
where
\begin{eqnarray}&&\nonumber
W_{\mu\nu}=-\frac{1}{6}g_{\mu\nu}R^{;\lambda}_{;\lambda}+\frac{2}{3}R_{;\mu;\nu}+R_{\mu\nu;\lambda}^{;\lambda}-R_{\lambda\nu;\mu}^{;\lambda}-R_{\lambda\mu;\nu}^{;\lambda}\\
&&+\frac{2}{3}RR_{\mu\nu}-2R_{\mu\lambda}R_\nu^\lambda+\frac{1}{2}g_{\mu\nu}R_{\lambda\kappa}R^{\lambda\kappa}-\frac{1}{6}g_{\mu\nu}R^2,
\end{eqnarray}
and the energy-momentum tensor of matter is
\begin{eqnarray}&&\nonumber
T^{\mu\nu}= i\overline{\psi}\gamma^\mu D^\nu \psi+\frac{2}{3}S^{;\mu}S^{;\nu}-\frac{1}{6}g^{\mu\nu}S^{;\kappa}S_{;\kappa}-\frac{1}{3}SS^{;\mu;\nu}\\
&&+\frac{1}{3}g^{\mu\nu}SS^{;\kappa}_{;\kappa}-g^{\mu\nu}\lambda S^4-\frac{1}{6}S^2\left(R^{\mu\nu}-\frac{1}{2}g^{\mu\nu}R\right).
\end{eqnarray}
By using local conformal invariance, the energy-momentum tensor can be written as
\begin{eqnarray}
T^{\mu\nu}=T^{\mu\nu}_{\rm kin}-\frac{1}{6}S_0^2\left(R^{\mu\nu}-\frac{1}{2}g^{\mu\nu}R\right)-g^{\mu\nu}\lambda S_0^4.
\end{eqnarray}
where $T^{\mu\nu}_{\rm kin}=i\overline{\psi}\gamma^\mu D^\nu \psi$ and $S_0$ is a constant. According to Eq. (\ref{field}) and $W^{\mu\nu}=0$, we have
\begin{eqnarray}
\label{field1}
T^{\mu\nu}_{\rm kin}-g^{\mu\nu}\lambda S_0^4=\frac{1}{6}S_0^2\left(R^{\mu\nu}-\frac{1}{2}g^{\mu\nu}R\right).
\end{eqnarray}

Considering a perfect fluid, $T^{\mu\nu}_{\rm kin}=(\rho+p)u^\mu u^\nu+pg^{\mu\nu}$, in the Friedmann-Robertson-Walker-Lema\^{i}tre (FRWL) spacetime with the scale
factor $a(t)$
\begin{eqnarray}
\label{frwmet}
ds^2=-dt^2+a^2(t)\left[\frac{dr^2}{1-Kr^2}+r^2(d\theta^2+\sin^2\theta
d\phi^2)\right],
\end{eqnarray}
where the spatial curvature constant $K=+1$, 0, and $-1$ corresponds
to a closed, flat and open Universe, respectively, Eq. (\ref{field1}) takes the form
\begin{eqnarray}
\label{feq}
H^2+\frac{K}{a^2}=-\frac{2\rho_{\rm m}}{S^2_0}-2\lambda S^2_0,
\end{eqnarray}
where $H=\dot{a}/a$ and $\rho_{\rm m}$ represents energy density of matter which can be separate into a relativistic and a non-relativistic component: $\rho_{\rm m}=\rho_{\rm nr}+\rho_{\rm r}=\rho_{\rm nr0}a^3+\rho_{\rm r0}a^4$. Taking substitutions $G=-3/(4\pi S^2_0)$ and $\Lambda=-6\lambda S^2_0$, Eq. (\ref{feq}) is identical to the standard Friedmann equation: $H^2+K/a^2=8\pi G\rho/3+\Lambda/3$. Both $G$ and $\Lambda$, however, are negative and depend on the same parameter $S^2_0$ in the oppositive way in CG. Constrained from the rotation curves of spiral galaxies, $K$ must be negative: $K<0$ \cite{Mannheim1997}.

If we define $\Theta_{\rm m}\equiv 2\rho_{\rm m}/(H^2S^2_0)=\Theta_{\rm nr}+\Theta_{\rm r}$, $\Theta_{\rm \Lambda}\equiv-2\lambda S^2_0/H^2$, and $\Theta_{\rm K}\equiv-K/(a^2H^2)$ (because $K<0$ and $\lambda<0$, all these parameter are positive), Eq. (\ref{feq}) yields: $\Theta_{\rm \Lambda}+\Theta_{\rm K}-\Theta_{\rm nr}-\Theta_{\rm r}=1$. By taking the derivative of Eq. (\ref{feq}), we obtain
\begin{eqnarray}
\label{feq1}
\frac{\ddot{a}}{a}=H^2 \left(\Theta_{\rm \Lambda}+\Theta_{\rm r}+\frac{\Theta_{\rm nr}}{2}\right),
\end{eqnarray}
which is always positive, whereas the deceleration parameter
\begin{eqnarray}
\label{de}
q\equiv -\frac{\ddot{a}a}{\dot{a}}=-\Theta_{\rm \Lambda}-\Theta_{\rm r}-\frac{\Theta_{\rm nr}}{2},
\end{eqnarray}
is always negative. So the expansion of the universe in CG accelerates at all times, unlike the standard cosmology. With the present values of $\Theta$ parameters, Eq. (\ref{feq}) can reexpressed as
\begin{eqnarray}
\label{feq2}
H^2=H_0^2 \left(\Theta_{\rm \Lambda 0}+\Theta_{\rm K0}a^{-2}-\Theta_{\rm nr 0}a^{-3}-\Theta_{\rm r 0}a^{-4}\right),
\end{eqnarray}
which is analogous to the standard Friedmann equation, but has negative signs in front of the matter parameters $\Theta_{\rm nr 0}$ and $\Theta_{\rm r 0}$. This equation implies that in CG $a$ will reach a minimum value $a_{\rm min}>0$, rather than the singularity $a=0$. By adjusting parameters, we can obtain smaller and smaller $a_{\rm min}$ (lager and lager $z_{\rm max}$). Because there is no need to introduce dark matter in CG, terms $\Theta_{\rm nr 0}$ and $\Theta_{\rm r 0}$ can be neglect for large scale factor $a$, comparing with terms $\Theta_{\rm \Lambda 0}$ and $\Theta_{\rm K0}$. The present age of the universe is found to be
\begin{eqnarray}
\label{age}
H_0t_0=\frac{1}{\sqrt{-q_0}} {\rm~atanh}(\sqrt{-q_0}).
\end{eqnarray}
\section{Observational constraints on conformal gravity}
In this section, we use the Union2.1 SNIa data and the observational Hubble parameter data to consider observational bounds on CG, and test it with the age of an old quasar by using the best-fit values constrained from SNIa and $H(t)$ data.
\subsection{Constrain conformal gravity with SNIa and $H(t)$ data}
The SNIa data provide the main evidence for the existence of dark energy in the framework of standard cosmology.
The Union2.1 compilation, consisting of 580 SNIa data \cite{Suzuki2012,Amanullah2010,Amanullah2008}, is the largest published and spectroscopically
confirmed SNIa sample to date. Each SNIa data
point at redshift $z_i$ includes the Hubble-parameter free distance
modulus $\mu_{\rm obs}(z_i)$ ($\equiv m_{\rm obs}-M$, where $M$ is
the absolute magnitude) which is derived from the direct observables provided by the SNIa data and is not a directly observable quantity and the corresponding error $\sigma^2_i$.
The resulting theoretical distance modulus $\mu_{\rm th}(z)$ is
defined as
\begin{eqnarray}
\label{SNIa}
\mu_{\rm th}(z)\equiv 5\log_{10}d_{\rm L}(z)+25,
\end{eqnarray}
where the luminosity distance in units of Mpc is expressed as \cite{Mannheim2003,Diaferio2011}
\begin{eqnarray}
\label{lu}
d_{\rm L}=\frac{(1+z)^2}{q_0H_0} \left[\left(1+q_0-\frac{q_0}{(1+z)^2}\right)^{1/2}-1\right].
\end{eqnarray}
This equation is obtained approximatively from equation (\ref{feq2}) and therefore is dependent on the cosmological model considered here, but it can be safely applied to the real Universe at sufficiently late times \cite{Diaferio2011}.

In order to discuss the age problem in CG in a consistent way, we treat $H_0$ as a parameter rather than marginalize it over in data-fitting. Assuming the measurement errors are Gaussian, the likelihood function is ${\cal{L}} \propto e^{-\chi^2/2}$. The
model parameters yield a minimal $\chi^{2}$ and a maximal
${\cal{L } }$ will be favored by the observations. The $\chi^{2}$ function for SNIa data is
\begin{eqnarray}
\chi^{2}_{\rm SNIa}(q_0, H_0)&=&\sum^{580}_{i=1}\frac{[\mu^{\rm obs}_{\rm L}(z_i)-\mu^{\rm th}_{\rm L}(z_i)]^2}
{\sigma^2_i}.
\end{eqnarray}

We also use 28 Hubble parameter data to constrain CG. Based on the work in \cite{Jimenez}, 9 $H(z)$ data were obtained by using the
age of evolving galaxies \cite{Simon}. These data were revised at 11 redshifts from the differential
ages of red-envelope galaxies \cite{Stern}. 2 $H(z)$ data were obtained by taking the BAO scale as a standard ruler in the radial direction \cite{Gaztanaga}. 3 $H(z)$ data were acquired by combining measurements of the
baryon acoustic peak and Alcock-Paczynski distortion from galaxy clustering in the WiggleZ Dark Energy Survey \cite{Blake}.
Recently, 8 $H(z)$ data were obtained from the differential spectroscopic evolution of
early-type galaxies as a function of redshift \cite{Moresco}. Other 4 $H(z)$ data were presented in \cite{Zhang}. Observed values of
the Hubble parameter can be used to constrain the parameters of CG. The $\chi^{2}$ function of the $H(z)$ data is given by
\begin{eqnarray}
\chi^{2}_{\rm H}(q_0, H_0)&=&\sum^{28}_{i=1}\frac{[H_{\rm obs}(z_i)-H_{\rm
th}(z_i)]^2}{\sigma^2_{{\rm H}_i}}.
\end{eqnarray}
Since the SNIa and $H(z)$ data are effectively independent measurements, we can minimize their total $\chi^{2}$ value given by
\begin{eqnarray}
\label{21}\chi^2(q_0, H_0)=\chi^{2}_{\rm SNIa}+\chi^{2}_{\rm H},
\end{eqnarray}
to find the best-fit values of the parameters of CG.

Constraining CG only with SNIa data, we find the best-fit values of the parameters at 68.3\% confidence as: $q_0=-0.29 \pm 0.07$ and $H_0=69.15\pm 0.57$ km s$^{-1}$Mpc$^{-1}$ with $\chi^2_{\rm min}=575.41$ ($\chi^2_{\rm min}/$dof=0.99, dof is the logogram of degree of freedom), as shown in Table \ref{tab2}. If 28 $H(z)$ data points are also included in fitting, we find the best-fit values of the parameters at 68.3\% confidence as: $q_0=-0.33 \pm 0.06$ and $H_0=69.3\pm 0.5$ km s$^{-1}$Mpc$^{-1}$ with $\chi^2_{\rm min}=599.9$ ($\chi^2_{\rm min}/$dof=0.99), also as shown in Table \ref{tab2}.

\begin{table*}
\caption{\label{tab2}The best-fit values of the parameters ($q_0$, $H_0$) of CG with the corresponding the formation redshift $z_{\rm f}$ and
$\chi^2_{\rm min}/$dof fitting from
SNIa and SNIa$+H(z)$ observations with 1 $\sigma$
confidence level, here $H_0$ with dimension km s$^{-1}$Mpc$^{-1}$.}
\begin{tabular}{c|c|c|c}\hline\hline
 Observations     & $q_0$   & $H_{0}$   & $\chi^2_{\rm min}/$dof    \\ \hline
 SNIa & $-0.29 \pm 0.07$  & $69.15\pm 0.57$  & $0.99$ \\ \hline
 SNIa$+H(z)$ &$-0.33 \pm 0.06$  &$69.3\pm 0.5$  & $0.99$   \\ \hline\hline
\end{tabular}
\end{table*}
In \cite{Mannheim2003}, Mannheim obtained $q_0=-0.37$ by constraining CG from 54 SNIa data. In \cite{Diaferio2011}, Diaferio et al. obtained the best-fit value as: $q_0=-0.12^{+0.08}_{-0.16}$ constrained from 115 GRBs; $q_0=-0.225^{+0.068}_{-0.066}$ constrained from 397 SNIa data; and $q_0=-0.164^{+0.015}_{-0.022}$ constrained from 115 GRBs and 397 SNIa data. In \cite{Kumar2012}, constraints on parameters at $1\sigma$ level in power-law cosmology were obtained as: $q_0=-0.18\pm 0.12$ and $H_0=68.4\pm 2.8$ kms$^{-1}$Mpc$^{-1}$ from $H(z)$ data, $q_0=-0.38\pm 0.05$ and $H_0=69.18\pm 0.55$ kms$^{-1}$Mpc$^{-1}$ from SNIa data, and $q_0=-0.34\pm 0.05$ and $H_0=68.93\pm 0.53$ kms$^{-1}$Mpc$^{-1}$ from the joint test using $H(z)$ and SNe Ia data. Recently, Planck 2013 results found a low value of the Hubble constant in the framework of $\Lambda$CDM model: $H_0=67.3\pm 1.2$ km s$^{-1}$Mpc$^{-1}$ \cite{Planck}. Our constraints on parameters of CG are consistent with all these results.

In \cite{Diaferio2011}, a Bayesian approach has been used to infer the cosmological parameters, while a $\chi^{2}$ procedure based on the distance modulus equation (\ref{SNIa}) is used here and in \cite{Mannheim2003}.	The distance modulus equation (\ref{SNIa}) is not a directly observable quantity but derives from assumptions on the cosmological model, in other words, the $\chi^{2}$ procedure is model-dependent, while the Bayesian approach does not. Strictly speaking, a Bayesian approach is required to constrain CG with observations, as done in \cite{Diaferio2011}. The difference between the Bayesian approach and the $\chi^{2}$ procedure may can explain the two to three $\sigma$ difference between the values of $q_0$ derived in \cite{Diaferio2011} and the values derived in \cite{Mannheim2003} or in our work. The degeneracies between different parameters may be another factor resulted to the different values of $q_0$ obtained from Bayesian approach or from $\chi^{2}$ procedure.

\subsection{Test conformal gravity with the age of old high-redshit objects}
Old high-redshit objects are usually used to constrain parameters or test cosmological models \cite{Lima2009}.
Recently, many dark energy models have been tested with the age of APM~08279+5255, such as $\Lambda$CDM \cite{Friaca2005, Alcaniz2003}, the creation of cold dark matter models \cite{Chen2012},
$\Lambda(t)$ model \cite{Cunha2004},
the new agegraphic dark energy \cite{Li2011}, parametrized variable Dark Energy
Models \cite{Barboza2008, Dantas2007}, the $f(R)=\sqrt{R^2-R^2_0}$ model \cite{Movahed2007}, quintessence \cite{Capozziello2007, Jesus2008}, holographic dark energy model \cite{Granda2009,Wei2007}, braneworld modes \cite{Movahed2007a, Pires2006, Movahed2008, Alam2006}, and other models \cite{Sethi2005, Abreu2009, Santos2008}, and it has been shown that none of these dark energy models can accommodate the age of quasar APM~08279+5255.

In order to investigate whether CG can accommodate the quasar APM~08279+5255, we must first understand the possible range of the age of
APM~08279+5255. From \emph{XMM-Newton} observations of APM~08279+5255, an iron overabundance of
Fe/O of $3.3\pm0.9$ (the abundance ratio has been normalized to the solar value) has been derived for the broad absorption line system \cite{Hasinger}, and the age of APM~08279+5255 was estimated to lie within the interval 2-3 Gyr by using an Fe/O=3 abundance ratio derived from X-ray observations \cite{Komossa2003}.
In \cite{Friaca2005}, the age of APM~08279+5255 was re-evaluated by using
a chemodynamical model for the evolution of spheroids: an age of 2.1
Gyr was obtained when the Fe/O abundance ratio of the model
reaches 3.3 which is the best-fit value acquired in \cite{Hasinger}; an age of 1.8 Gyr was obtained when the Fe/O abundance ratio
reaches 2.4 which is 1 $\sigma$ deviation from the best-fitting value; an age of 1.5 Gyr was set when the Fe/O abundance ratio reaches 2 which is a highly improbable value of Fe/O, as it
would require $N_{\rm H}$ in excess of $1.2\times10^{23}$ cm$^{-2}$
(see Fig.~3 in \cite{Hasinger}), which seems to be ruled
out from the determinations of $N_{\rm H}=(5.3-9.1)\times10^{22}$
cm$^{-2}$ by other \emph{Chandra} and \emph{XMM} observations; even
considering only the \emph{XMM}2 data set, the lowest value of Fe/O
is $2.4$  at $N_{\rm H}=1.28\times10^{22}$ cm$^{-2}$ (see Fig.~3 in \cite{Hasinger}).
So the age of APM~08279+5255 since the initial star formation has been estimated as \cite{Yang2010}: (1) the best estimated value is 2.1 Gyr; (2) 1 $\sigma$ lower limit is 1.8 Gyr; (3) the lowest limit
is 1.5 Gyr.

The age of a cosmic object (for example, a quasar, a galaxy, or a galaxy cluster) is defined as the difference between the age of the Universe at redshift $z$ and
the one when the object was born (at its formation redshift $z_{\rm f}$)
\begin{eqnarray}
T(z)=\int_{z}^{z_{\rm f}} \frac{dz'}{(1+z')H(z')}.
\end{eqnarray}
To test CG with the age of APM~08279+5255, we must also know the formation redshift $z_{\rm f}$ of APM~08279+5255. This redshift, however, can only be inferred from available
observational results. In many literatures, the $z_{\rm f}$ of APM~08279+5255 have been taken as infinity, which is incorrect. Only in \cite{Yang2010}, a finite $z_{\rm f}$ has been taken into account. Following \cite{Yang2010}, we will take a finite $z_{\rm f}$ to discuss the age problem in CG. Unlike in \cite{Yang2010} the $z_{\rm f}$ of APM~08279+5255 in the frame of $\Lambda$CDM can be inferred from WMAP data: the peak epoch of reionizing was found to be at $z_{\rm reion}=17 \pm 10$ from WMAP1 data \cite{Bennett} and
at $z_{\rm reion}=10.8 \pm 1.4$ from WMAP5 data \cite{Dunkley}, and the star formation processes could be inferred as early as $z=15-17$ in high density peaks. The $z_{\rm f}$ of APM~08279+5255 in CG, however, should be inferred from observations which is in dependent on cosmological models. Results based on the new Hubble WFC3/IR imaging in the
Ultradeep Field implied that the global star formation rate density might start from a very high value at $z\approx 10$ \cite{Yan}. Take into account this result and take the results from WMAP data for reference, we take $z_{\rm f}=15$ as the formation redshift of APM~08279+5255 in the discussions.

\begin{figure}
\includegraphics[width=10cm]{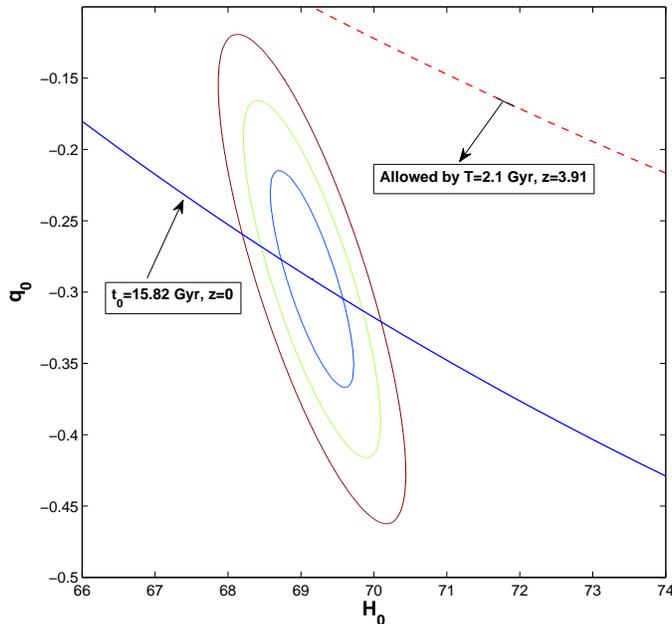}
\caption{The $68.3\%$, $95.4\%$ and $99.7\%$ confidence regions in the $H_0$-$q_{0}$ ($H_0$ with dimension kms$^{-1}$Mpc$^{-1}$) plane fitting from SNIa data. The line with $t_0=15.82$ Gyr at $z=0$ and the region allowed by the age of APM~08279+5255 are also shown. \label{Figs}}
\end{figure}

\begin{figure}
\includegraphics[width=10cm]{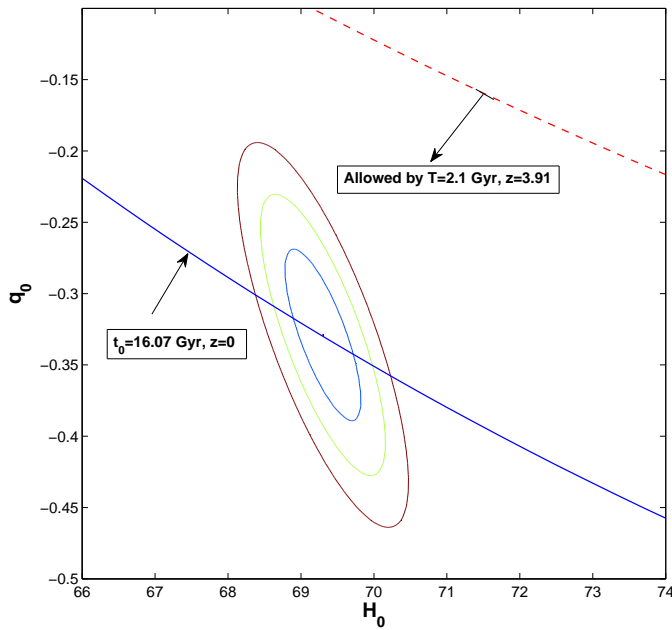}
\caption{The $68.3\%$, $95.4\%$ and $99.7\%$ confidence regions in the $H_0$-$q_{0}$ ($H_0$ with dimension kms$^{-1}$Mpc$^{-1}$) plane fitting from SNIa and $H(t)$ data. The line with $t_0=16.07$ Gyr at $z=0$ and the region allowed by the age of APM~08279+5255 is also shown. \label{Figsh}}
\end{figure}

Taking $q_0=-0.29$ and $H_{\rm 0}=69.15$ km
s$^{-1}$Mpc$^{-1}$ fitted from SNIa data only, and according to Eq.(\ref{age}), we find the present age of the Universe in CG is $t_0=15.82$
Gyr, larger than 14 Gyr estimated from old globular clusters \cite{Pont}. CG accommodates the age of APM~08279+5255 at 3 $\sigma$ deviation, as shown in Figure \ref{Figs}. Taking $q_0=-0.33$ and $H_{\rm 0}=69.3$ kms$^{-1}$Mpc$^{-1}$ fitted from SNIa and $H(t)$ data, we find the present age of the Universe in CG is $t_0=16.07$
Gyr, also larger than 14 Gyr estimated from old globular clusters \cite{Pont}. In this case, CG also accommodates the age of APM~08279+5255 at 3 $\sigma$ deviation, as shown in Figure \ref{Figsh}.

\section{Conclusions and discussions}
We have constrained the conformal gravity with 580 SNIa data and 28 $H(t)$ data. We obtained the best-fit values of the parameter at $68\%$
confidence level: $q_0=-0.29 \pm 0.07$ and $H_0=69.15\pm 0.57$ kms$^{-1}$Mpc$^{-1}$ constrained from 580 SNIa data; and $q_0=-0.33 \pm 0.06$ and $H_0=69.3\pm 0.5$ kms$^{-1}$Mpc$^{-1}$ constrained from 580 SNIa data and 28 $H(t)$ data. With these best-fit
values of the parameters of CG, we test CG with the age of APM~08279+5255. We have found that CG can accommodate the best estimated value of
the age of APM~08279+5255 at 3 $\sigma$ deviation.

In the discussions, we have ignored the contributions of radiation and non-relativistic matter. If their contributions are take into account, the age of APM~08279+5255 will be a little lager. In this cases, CG can more easily accommodate the best estimated value of the age of APM~08279+5255. We can conclude that unlike most of dark energy models CG does not suffer from a problem with the estimated age of APM~08279+5255 at redshift $z=3.91$, based on the values of
Hubble constant $H_0$ and the deceleration parameter $q_{0}$ constrained from the currently available SNIa and $H(z)$ data. The results we obtained here can be tested with future cosmological observations.

\begin{acknowledgments}
This study is supported in part by National Natural Science Foundation of China (Grant Nos. 11147028, 11273010, and 11103019) and Hebei Provincial Natural Science Foundation of China (Grant No. A2011201147).
\end{acknowledgments}

\bibliography{apssamp}

\begin{thebibliography}{99}

\bibitem{weinberg}
S. Weinberg, Rev. Mod. Phys. 61(1989) 1; S. M. Carroll, Living Rev. Rel. 4 (2001) 1.

\bibitem{Yang2010}
R.-J. Yang, S. N. Zhang, Mon. Not. R. Astron. Soc. 407 (2010) 1835.

\bibitem{Felice}
A. D. Felice, S. Tsujikawa, Living Rev. Rel. 13 (2010) 3.

\bibitem{Sotiriou}
T. P. Sotiriou, V. Faraoni, Rev. Mod. Phys. 82 (2010) 451.


\bibitem{Capozziello}
K. Bamba, S. Capozziello, S. Nojiri, S. D. Odintsov, Astrophys. Space Sci. 342 (2012) 155.


\bibitem{Clifton2012}
T. Clifton, P. G. Ferreira, A. Padilla, C. Skordis, Phys. Rep. 513 (2012) 1


\bibitem{Bengochea}
G. R. Bengochea, R. Ferraro, Phys. Rev. D 79 (2009) 124019.

\bibitem{Yang2011}
R. Yang, Eur. Phys. J. C 71 (2011) 1797.

\bibitem{Carroll}
S. M. Carroll, A. D. Felice, V. Duvvuri, D. A. Easson, M. Trodden, M. S. Turner, Phys. Rev. D 71 (2005) 063513.

\bibitem{Clifton}
T. Clifton, J. D. Barrow, Phys. Rev. D 72 (2005) 123003.

\bibitem{Navarro}
I. Navarro, K. Van Acoleyen, J. Cosmol. Astropart. Phys. 03 (2006) 008.

\bibitem{Weyl1918}
H. Weyl, Reine Infinitesimalgeometrie, Math. Z. 2 (1918) 384.

\bibitem{Nesbet}
R. K. Nesbet, Entropy 15 (2013) 162.

\bibitem{Scholz}
 E. Scholz, arXiv:1111.3220

\bibitem{Mannheim2011}
P. D. Mannheim, arXiv:1101.2186

\bibitem{Mannheim2001}
P. D. Mannheim, Prog. Part. Nucl. Phys. 56 (2006) 340; Astrophys. J. 561 (2001) 1.

\bibitem{Mannheim1997}
P. D. Mannheim, Astrophys. J. 479 (1997) 659; J. G. O'Brien, P. D. Mannheim, Mon. Not. R. Astron. Soc. 421 (2012) 18351273.

\bibitem{Said}
J. L. Said, J. Sultana, K. Z. Adami, Phys. Rev. D 86 (2012) 104009.

\bibitem{Mannheim2012}
P. D. Mannheim, Phys. Rev. D 85 (2012) 124008.

\bibitem{Mannheim2003}
P. D. Mannheim, Int. J. Mod. Phys. D12 (2003) 893.

\bibitem{Diaferio2011}
 A. Diaferio, L. Ostorero, V. F. Cardone, JCAP10(2011)008


 \bibitem{Flanagan}
E. E. Flanagan, Phys. Rev. D 74 (2006) 023002.

 \bibitem{Cattani2013}
 C. Cattani, M. Scalia, E. Laserra, I. Bochicchio, K. K. Nandi, Phys. Rev. D 87 (2013) 047503.

\bibitem{Pireaux}
S. Pireaux, Class. Quantum Grav. 21 (2004) 1897.

\bibitem{Horne}
K.Horne, Mon. Not. Roy. Astron. Soc. 369 (2006) 1667



 \bibitem{Diaferio}
A. Diaferio, L. Ostorero, Dip. F. Generale, Mon Not R Astron Soc 393 (2009) 215.

\bibitem{Elizondo}
 D. Elizondo and G. Yepes, Astroph. J. 428 (1994) 17.

\bibitem{Pont}
 F. Pont, M. Mayor, C. Turon, D. A. Vandenberg, A\&A 329 ( 1998) 87.

\bibitem{Komossa2003}
 S. Komossa, G. Hasinger, 2003, in Hasinger G., Baller Th., Parmar A. N., eds, XEUS studying the evolution of the universe, MPE Report.
MPE, Garching, P.281

\bibitem{Friaca2005}
 A. Friaca, J. Alcaniz, J. A. S. Lima,Mon. Not. R. Astron. Soc. 362 (2005) 1295.

 \bibitem{Alcaniz2003}
 J. S. Alcaniz, J. A. S. Lima, J. V. Cunha, Mon. Not. R. Astron. Soc. 340 (2003) L39.

\bibitem{Chen2012}
J. Chen, P. Wu, H. Yu, Z. Li, Eur. Phys. J. C 72 (2012) 1861.

\bibitem{Cunha2004}
 J. V. Cunha, R. C. Santos, Int. J. Mod. Phys. D 13 (2004) 1321.

\bibitem{Li2011}
Y.-H. Li, J.-Z. Ma, J.-L. Cui, Z. Wang, X. Zhang, Sci.China Phys. Mech. Astron.54 (2011) 1367.

\bibitem{Barboza2008}
 E. M. Jr. Barboza, J. S. Alcaniz, Phys. Lett. B 666 (2008) 415.

\bibitem{Dantas2007}
 M. A. Dantas, J. S. Alcaniz, D. Jain, A. Dev, Astron. Astrophys. 467 (2007) 421.


\bibitem{Movahed2007}
 M. S. Movahed, S. Baghram, S. Rahvar, Phys. Rev. D 76 (2007) 044008.

\bibitem{Capozziello2007}
 S. Capozziello, P. K. S. Dunsby, E. Piedipalumbo, C. Rubano, Astron. Astrophys. 472 (2007) 51.

\bibitem{Jesus2008}
 J. F. Jesus, R. C. Santos, J. S. Alcaniz, J. A. S. Lima, Phys. Rev. D 78 (2008) 063514.

\bibitem{Granda2009}
 L. N. Granda, A. Oliveros, W. Cardona, arXiv:0905.1976.

\bibitem{Wei2007}
 H. Wei, S. N. Zhang, Phys. Rev. D. 76 (2007) 063003.

\bibitem{Movahed2007a}
 M. S. Movahed, S. Ghassemi, Phys. Rev. D 76 (2007) 084037.

\bibitem{Movahed2008}
 M. S. Movahed, A. Sheykhi, Mon. Not. R. Astron. Soc. 388 (2008) 197.

\bibitem{Alam2006}
 U. Alam, V. Sahni, Phys. Rev. D 73 (2006) 084024.

\bibitem{Pires2006}
 N. Pires, Z. H. Zhu, J. S. Alcaniz, Phys. Rev. D 73 (2006) 123530.

\bibitem{Santos2008}
 R. C. Santos, J. V. Cunha, J. A. S. Lima, Phys. Rev. D 77 (2008) 023519.

\bibitem{Sethi2005}
 G. Sethi, A. Dev, D. Jain, Phys. Lett. B 624 (2005) 135.

\bibitem{Abreu2009}
 E. M. C. Abreu, L. P. G. De Assis, C. M. L. dos Reis, Int. J. Mod. Phys. A 24 (2009) 5427.

\bibitem{Bender2008}
C. M. Bender and P. D. Mannheim, Phys. Rev. Lett. 100 (2008) 110402; C. M. Bender and P. D. Mannheim, Phys. Rev. D 78 (2008) 025022.

\bibitem{Mannheim1992}
P. D. Mannheim, Astroph. J. 391 (1992) 429.

\bibitem{Suzuki2012}
N. Suzuki et al., Astroph. J. 746 (2012) 85.

\bibitem{Amanullah2010}
R. Amanullah et al. Astrophys.J.716 (2010) 712.

\bibitem{Amanullah2008}
M. Kowalski, Astrophys. J. 686 (2008) 749.

\bibitem{Jimenez}
R. Jimenez, L. Verde, T. Treu, D. Stern, Astroph. J. 593 (2003) 622.

\bibitem{Simon}
J. Simon, et al., Phys. Rev. D 71 (2005) 123001.

\bibitem{Stern}
D. Stern, R. Jimenez, L. Verde, M. Kamionkowski, S. Adam Stanford, J. Cosmol. Astropart. Phys. 02 (2010) 008.


\bibitem{Gaztanaga}
 E. Gaztanaga, A. Cabr¡äe, L. Hui. Mon. Not. R. Astron. Soc. 399 (2009) 1663.

\bibitem{Blake}
C. Blake et al., Mon. Not. R. Astron. Soc. 425 (2012) 405.

\bibitem{Moresco}
M. Moresco et al., J. Cosmol. Astropart. Phys. 08 (2012) 006.

\bibitem{Zhang}
C. Zhang, H. Zhang, S. Yuan, T.-J. Zhang, Y.-C. Sun, arXiv:1207.4541.



\bibitem{Lima2009}
J. A. S. Lima, J. F. Jesus, J. V. Cunha, Astrophys. J. Lett. 690 (2009) L85.

\bibitem{Kumar2012}
S. Kumar, Mon. Not. R. Astron. Soc. 422 (2012) 2532-2538.

\bibitem{Planck}
Planck Collaboration: P. A. R. Ade et al., arXiv:1303.5076

\bibitem{Hasinger}
 G. Hasinger, N. Schartel, and S. Komossa, Astrophys. J. 573 (2002) L77.

 \bibitem{Bennett}
 C. L. Bennett et al., Astrophys.J.Suppl.148 (2003) 1.

 \bibitem{Dunkley}
 J. Dunkley et al., Astrophys. J. Suppl. 180 (2009) 306.

\bibitem{Yan}
H.-J. Yan et al., Res. Astron. Astrophys. 10 (2010) 867.

\end{thebibliography}

\end{document}